\documentclass[sigconf, screen]{acmart}
\acmConference[ESEC/FSE 2023]{The 31st ACM Joint European Software Engineering Conference and Symposium on the Foundations of Software Engineering}{11 - 17 November 2023}{San Francisco, USA}
\usepackage{orcidlink}
\usepackage{multirow}
\usepackage{microtype}
 
\newif\ifSPACEHACK
\SPACEHACKtrue 

\newif\ifDEBUG
\DEBUGtrue

\newif\ifANONYMOUS
\ANONYMOUSfalse

\usepackage{amsmath,amssymb,amsfonts}
\usepackage{algorithmic}
\usepackage{graphicx}
\usepackage{textcomp}
\usepackage{xcolor}
\usepackage{booktabs} 
\usepackage{xspace} 
\usepackage[normalem]{ulem}
\usepackage{makecell}
\usepackage{tcolorbox}
\usepackage{enumitem}
\usepackage{siunitx}
\usepackage[vskip=1em,font=itshape,leftmargin=2em,rightmargin=2em]{quoting}
\usepackage{hyperref}

\usepackage[noadjust]{cite}

\usepackage{soul}
\usepackage{multirow}
\usepackage{array,booktabs}
\newcolumntype{M}[1]{>{\centering\arraybackslash}m{#1}}
\usepackage{amsfonts}
\usepackage{amsmath}
\usepackage{wrapfig, lipsum}
\usepackage{subfigure}

\ifDEBUG
    \newcommand{\JD}[1]{\textcolor{purple}{[JD:#1]}}
    \newcommand{\AG}[1]{\textcolor{olive}{[AG:#1]}}
    \newcommand{\WJ}[1]{\textcolor{olive}{[WJ:#1]}}
    \newcommand{\GKT}[1]{\textcolor{brown}{[GKT:#1]}}
    \newcommand{\MH}[1]{\textcolor{pink}{[MH:#1]}}
    \newcommand{\NMS}[1]{\textcolor{orange}{[NMS:#1]}}
    \newcommand{\WPM}[1]{\textcolor{red}{[Trey: #1]}}
    \newcommand{\TRS}[1]{\textcolor{teal}{[Taylor: #1]}}
    \newcommand{\Kelechi}[1]{\textcolor{olive}{[Kelechi: #1]}}
    
    \newcommand{\TODO}[1]{\hl{#1}}
\else
    \newcommand{\JD}[1]{}
    \newcommand{\AG}[1]{}
    \newcommand{\WJ}[1]{}
    \newcommand{\GKT}[1]{}
    \newcommand{\NS}[1]{}
    \newcommand{\NV}[1]{}
    \newcommand{\NMS}[1]{}
    \newcommand{\WPM}[1]{}
    \newcommand{\TRS}[1]{}
    \newcommand{\MH}[1]{}
    
    \newcommand{\TODO}[1]{}
\fi

\newif\ifSPACEHACK
\SPACEHACKfalse

\newcommand{\myparagraph}[1]{\vspace{0.07cm}\noindent\ul{\emph{\textbf{#1}}}\noindent{}}

\ifSPACEHACK
    \titlespacing*\section{0pt}{4pt plus 2pt minus 2pt}{4pt plus 2pt minus 2pt}
    \titlespacing*\subsection{0pt}{4pt plus 1.5pt minus 1.5pt}{4pt plus 1.5pt minus 1.5pt}
    \titlespacing*\subsubsection{0pt}{3pt plus 1.5pt minus 1.5pt}{3pt plus 1.5pt minus 1.5pt}
    \titlespacing*\paragraph{0pt}{2pt plus 1.5pt minus 1.5pt}{2pt plus 1.5pt minus 1.5pt}
\fi
\usepackage{balance} 

\usepackage{amsmath}
\usepackage{cleveref}

\crefformat{section}{\S#2#1#3}
\crefname{figure}{Figure}{Figures}
\crefname{appendix}{Appendix}{Appendices}
\crefname{table}{Table}{Tables}
\crefname{algorithm}{Algorithm}{Algorithms}
\crefname{listing}{Listing}{Listings}
\crefname{theorem}{Theorem}{Theorems}
\crefname{thm}{Theorem}{Theorems}
\crefname{lemma}{Lemma}{Lemmata}
\crefname{equation}{Eqt.}{Eqts.}
\crefformat{Grammar}{Grammar #1}

\usepackage{enumitem}
\usepackage{booktabs}
\usepackage{svg}
\usepackage{listings}

\newcommand{\eg}{\textit{e.g.,} }
\newcommand{\etal}{\textit{et al.}\xspace}

\copyrightyear{2023}
\acmYear{2023}
\setcopyright{rightsretained}
\acmConference[ESEC/FSE '23]{Proceedings of the 31st ACM Joint European Software Engineering Conference and Symposium on the Foundations of Software Engineering}{December 3--9, 2023}{San Francisco, CA, USA}
\acmBooktitle{Proceedings of the 31st ACM Joint European Software Engineering Conference and Symposium on the Foundations of Software Engineering (ESEC/FSE '23), December 3--9, 2023, San Francisco, CA, USA}
\acmDOI{10.1145/3611643.3613075}
\acmISBN{979-8-4007-0327-0/23/12}

\begin{document}

\newcommand{\YearOfSamples}{{2021,2022}\xspace}
\newcommand{\DateOfSamples}{{two}\xspace}

\newcommand{\SamplePopulation}{{65}\xspace}
\newcommand{\RandomSamplePopulation}{{33}\xspace}
\newcommand{\SelectedVenues}{{ICSE, ESEC/FSE, and ASE}\xspace}
\newcommand{\SelectedVenuesPopulation}{{3}\xspace}
\newcommand{\PilotPapers}{{13}\xspace}

\newcommand{\RoundsOfDiscussion}{{\TODO{3}}\xspace}

\newcommand{\KappaScoreConstruct}{{0.86}\xspace}
\newcommand{\KappaScoreRelationship}{{0.67}\xspace}
\newcommand{\KappaScoreDataSource}{{0.88}\xspace}

\newcommand{\SilentPapers}{{16}\xspace}
\newcommand{\DescriptivePapers}{{6}\xspace}
\newcommand{\ExperimentalPapers}{{11}\xspace}

\newcommand{\SilentPercent}{{49\%}\xspace}
\newcommand{\DescriptivePercent}{{18\%}\xspace}
\newcommand{\ExperimentalPercent}{{33\%}\xspace}

\newcommand{\PublicPercent}{{64\%}\xspace}
\newcommand{\PrivatePercent}{{24\%}\xspace}
\newcommand{\BothDataPercent}{{12\%}\xspace}

\newcommand{\ProcessPolicyPapers}{{11}\xspace}
\newcommand{\BothPapers}{{10}\xspace}
\newcommand{\ProductPapers}{{12}\xspace}
\newcommand{\SingleConstructPapers}{{23}\xspace}

\newcommand{\Public}{{21}\xspace}
\newcommand{\Private}{{8}\xspace}
\newcommand{\Both}{{4}\xspace}

\newcommand{\ProcessPolicyPercent}{{33\%}\xspace}
\newcommand{\BothPercent}{{30\%}\xspace}
\newcommand{\ProductPercent}{{37\%}\xspace}
\newcommand{\SingleConstructPercent}{{70\%}\xspace}

\newcommand{\mytitle}{}

\renewcommand{\mytitle}{Representation Of Engineering Teams Policies in Software Engineering Research}
\renewcommand{\mytitle}{Does Empirical Software Engineering Consider Policy and Process?}
\renewcommand{\mytitle}{Reflection: Empirical Software Engineering Must Consider Policy and Process}
\renewcommand{\mytitle}{Reflection: Policy-Process-Product Relationships in Empirical Studies}
\renewcommand{\mytitle}{Reflection: Empirical Studies of the Process-Product Relationship}
\renewcommand{\mytitle}{Reflecting on the use of the Policy-Process-Product Theory in Empirical Software Engineering}
\renewcommand{\mytitle}{Reflecting on the Use of the Policy-Process-Product Theory in Empirical Software Engineering}
\title{\mytitle}

\author{Kelechi G. Kalu}
\orcid{0000-0002-8749-9697}
\affiliation{%
    \institution{Purdue University, IN, USA}
    \country{}
}
\email{kalu@purdue.edu}

\author{Taylor R. Schorlemmer}
\orcid{0000-0003-2181-5527}
\affiliation{%
    \institution{Purdue University, IN, USA}
    \country{}
}
\email{tschorle@purdue.edu}

\author{Sophie Chen}
\orcid{0009-0000-4133-4910}
\affiliation{%
    \institution{University of Michigan, MI, USA}
    \country{}
}
\email{sophie.cy.chen@gmail.com}

\author{Kyle Robinson}
\orcid{0009-0004-6365-6645}
\affiliation{%
    \institution{Purdue University, IN, USA}
    \country{}
}
\email{robin489@purdue.edu}

\author{Erik Kocinare}
\orcid{0009-0007-9151-5008}
\affiliation{%
    \institution{Purdue University, IN, USA}
    \country{}
}
\email{ekocinar@purdue.edu}

\author{James C. Davis}
\orcid{0000-0003-2495-686X}
\affiliation{%
    \institution{Purdue University, IN, USA}
    \country{}
}
\email{davisjam@purdue.edu}

\begin{abstract} \label{sec: abstract}
The primary theory of software engineering is that an organization's Policies and Processes influence the quality of its Products.
We call this the \emph{PPP Theory}.
Although empirical software engineering research has grown common, it is unclear whether researchers are trying to evaluate the PPP Theory.
To assess this, we analyzed half (\RandomSamplePopulation) of the empirical works published over the last two years in three prominent software engineering conferences.
In this sample, \SingleConstructPercent focus on policies/processes or products, not both.
Only 33\% provided measurements relating policy/process and products.
We make four recommendations:
  (1) Use PPP Theory in study design;
  (2) Study feedback relationships;
  (3) Diversify the studied feed-forward relationships;
  and
  (4) Disentangle policy and process.
Let us remember that research results are in the context of, and with respect to, the relationship between software products, processes, and policies. 

\end{abstract}





\maketitle




\section{Introduction} 
\label{Intro}

\begin{figure}
    \centering
    \includegraphics[width=0.95\linewidth]{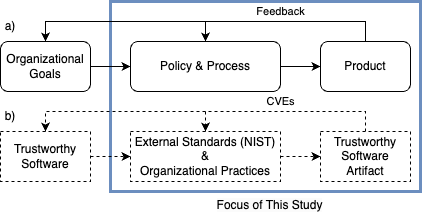}
    \caption{(a) Policy-Process-Product (PPP) Theory. Organizational goals influence the policies and processes adopted by software engineers. Policies and process influence product development. 
    Feedback may modify policies, processes, or the original goals.
    We treat the (often overlapped) concepts of Policy and Process as a single entity.
    (b) Example of the PPP Theory for the goal of producing trustworthy software. 
    }
    \label{fig: PPP model}
    \vspace{-0.25cm}
\end{figure}



Empirical software engineering research analyzes data to improve software products and engineering processes~\citep{basili_role_2006, xu_empirical_2017}. 
International standards organizations~\citep{ISO9001},
industry consortia~\citep{MISRA2012},
and professional organizations~\citep{IEEE730}
all assert that the \emph{Policies} and \emph{Processes} of software engineering influence the quality of the software \emph{Product} (the \emph{PPP Theory}).
Various studies support some of the relationships predicted by the PPP Theory~\citep{sommerville_software_2011, Petersen_context}.
Nevertheless, it remains unclear which policies and processes are most effective in achieving high-quality products, and how these vary by context~\citep{HobbsChapter3}.
To address this, experts have recommended that empirical software engineering researchers incorporate the PPP Theory, either as contextual information in case studies or as part of a controlled experiment~\citep{kitchenham_preliminary_2002, basili_building_1999}.
Doing so could resolve widely remarked-upon challenges with the generalizability and replicability of empirical software engineering research results~\citep{Petersen_context, dit_supporting_2013, nagappan_diversity_2013, madeyski_would_2017, murphy-hill_understanding_2010, siegmund_views_2015, kitchenham_preliminary_2002, gonzalez-barahona_reproducibility_2012}.
However, the extent to which the research community has taken this advice is unclear.

This reflection paper examines whether empirical software engineering researchers are considering the relationship between policies, processes, and software products. 
To achieve this, we reviewed empirical software research works published in \SelectedVenuesPopulation software engineering venues (\SelectedVenues) in 2021 and 2022.
We identified the primary aspects of the PPP Theory considered by each work, and the extent to which the PPP Theory was incorporated into the work.
We report that empirical studies consider a subset of the PPP Theory and are usually focused on individual theoretical concepts rather than the relationships of the theory.
We challenge the Empirical Software Engineering research community to consciously consider the PPP Theory in their study designs.



\section{Background: The PPP Theory} 
\label{sec: Background}

\subsection{Theoretical Constructs}


\textbf{Policy:}
 Policy has many meanings,
   including processes, artifacts, discourses, and bodies of knowledge about a field~\citep{colebatch_introduction_2018, cram2017organizational, ball_what_2015}.
 In the software engineering literature, policy means
    both 
    organizational strategies~\citep{noauthor_nist_nodate, sethi_using_1995, kafali_how_2017},
    and
    technical system behaviors~\citep{sloman_policy_1994, naldurg_modeling_2003, dulay_policy_2001}.
For PPP Theory, we define \textbf{policy} as
    \emph{an official statement of an organization's software engineering practices, derived from the organization's goals}. 

\vspace{0.1cm}
\noindent
\textbf{Process:}
A process consists of the steps followed to accomplish a task, \eg performing code review or implementing a new feature~\citep{noauthor_nist_nodate, sommerville_software_2011}.
For PPP Theory, we define \textbf{process} as \emph{the methods used by software engineers to accomplish their tasks}.

In the software engineering literature, we found that \textit{process} and \textit{policy} typically have overlapping definitions.
We lump them together into a single \textbf{process/policy} construct as shown in \cref{fig: PPP model}.

\vspace{0.1cm}
\noindent
\textbf{Product:}
A software product is a set of software and associated documentation, designed and developed to meet a specific set of user needs ~\citep{sommerville_software_2011, pressman2014software, iso/iec12207:2017}.
For PPP Theory, we define a \textbf{product} as \emph{the artifacts produced by a software engineering process}.
What comprises a product is context-dependent; some teams produce libraries, others web services, others mobile applications, and so on.

\subsection{Policy-Process-Product Relationship}

\cref{fig: PPP model} shows the PPP Theory: these constructs and the relationships between them.
Organizational goals are iteratively refined into policies, processes, and finally products.
This theory is propounded by documents from
 international standards organizations~\citep{ISO9001},
 industry consortia~\citep{MISRA2012},
 professional organizations~\citep{IEEE730},
 governments~\citep{noauthor_nist_nodate},
 and the academic literature~\citep{sloman_policy_1994,sethi_using_1995, Petersen_context,kafali_how_2017,sommerville_software_2011,anandayuvaraj2022reflecting,amusuo_reflections_2022}.

The PPP Theory predicts bi-directional relationships between the three constructs.
A software team's policy informs how its processes are defined, and a team's process influences the quality of the product.
In the reverse direction, retrospectives and postmortems provide feedback to modify processes and policies.



An example of the PPP Theory is demonstrated in~\cref{fig: PPP model}(b).
An organization has the goal of securing its artifact's supply chain ~\citep{okafor2022sok}.
Organizational leaders create a policy: ``Follow NIST standards''~\citep{Souppaya_Scarfone_Dodson_2022}.
Engineering teams comply through several process elements, such as
  code review (for code vulnerability inspection)
  and 
  using provenance certification tools (\eg Sigstore~\citep{newman2022sigstore}).
The desired product quality, a secure supply chain, is assessed: defects (\eg CVEs) provide feedback to improve the process.

Some seminal works explore the relationships between the PPP theory constructs~\citep{Humphrey_1996, Wohlin_Höst_Wesslén, Wohlin_Wesslen_1998, Wohlin_Runeson_Höst_Ohlsson_Regnell_Wesslén_2012}.
For example, Humphrey \etal~\citep{Humphrey_1996} and Wohlin \etal~\citep{Wohlin_Höst_Wesslén} demonstrated the impact of the Personal Software Process (PSP) on the software product (forward direction). In a follow-up study, Wohlin \etal showed software defects can be utilized in the Feedback direction to improve the PSP~\citep{Wohlin_Wesslen_1998}.

\section{Question and Methods}

\begin{figure*}
    \centering
    \includegraphics[width=0.80
    \linewidth]{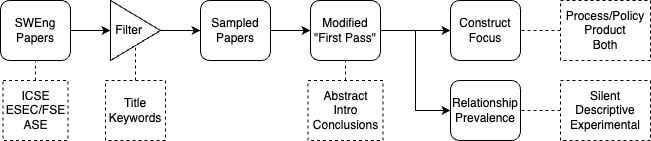}
    \caption{
    Our streamlined paper analysis approach. We begin by searching for full-length papers in leading software engineering conferences, filtering based on titles and keywords, resulting in empirical studies. We randomly sample half of these studies. This is followed by a "First Pass" analysis akin to Keshav ~\citep{howtoreadpaper} to comprehend paper content. We identify the paper's PPP Theory construct(s) focus and assess the presence of PPP Theory relationships.
    }
    \label{fig:method}
\end{figure*}


We ask:
\emph{To what extent does the PPP Theory inform modern empirical software engineering research?}
To answer this question, we assessed \RandomSamplePopulation papers from top software engineering research venues.
This section describes the selection of those papers, the initial assessment approach used in our pilot study, and our revised assessment approach.
Our final methodology is summarized in \cref{fig:method}.

\subsection{Paper selection}

We gathered recent empirical software engineering papers (2021-2022) from all tracks of three prominent conferences (ICSE, ESEC/FSE, and ASE).
We retrieved full-length papers, totaling 65, that included the term "empirical" in their title or keywords. Initially, we used the DBLP database for the title match and later cross-verified our findings and expanded our search using the ACM digital library, considering both the title and author's keywords. For analysis, we randomly selected 50\% of the collected papers.
\subsection{Analysis process}
Our goal was to assess the presence of PPP relationships in our selected papers.
We iteratively refined an analysis instrument through a pilot study. 
Ultimately, we assessed two distinct aspects of each work:
  its \textit{construct focus}
  and
  its \textit{relationship prevalence}.

\subsubsection{Pilot study}
In our pilot study, we established a complex classification scheme to rate the appearance of PPP Theory in empirical research papers. 
This includes a set of rating metrics and a method for scoring each paper on those metrics.

\myparagraph{Metrics:}
Our initial metrics attempted to measure the occurrence of process-product and policy-product relationships in each paper.
Each of these section-relationship combinations was evaluated on a four-point scale:
  (1) \textit{Silent} (no mention of relationships),
  (2) \textit{Implicit} (acknowledges relationship without discussion of impact),
  (3) \textit{Descriptive} (describes extensively the relationship between process/policy/product),
  and
  (4) \textit{Experimental} (describes and controls for these relationships in their experiment).

\myparagraph{Analysis Process:}
In our initial approach, raters focused on the Methodology, Results/Discussion, and Threats to Validity sections --- we thought any use of PPP Theory would be documented here.
After reading through a paper, raters classified the section-relationship combinations according to our four-point scale.

\myparagraph{Refinements:}
We used this approach on \PilotPapers papers in our pilot study (20\% of the available data).
We identified two flaws.
(1) Raters struggled to differentiate between levels of our four-point scale. Inter-rater disagreements were common and hard to resolve.
(2) Some PPP Theory elements were missed because the paper sections targeted in our analysis were too specific. 

\subsubsection{Final Analysis Approach}
First, we clarified definitions to make categories easier to differentiate.
Second, we characterized papers holistically rather than considering individual sections.
Lastly, given the relatively rare use of PPP Theory relationships in the pilot papers, we reduced the scope of the measurement to simply reporting whether process/product relationships were considered at all, or actively controlled for in the papers.
\cref{fig:method} provides an overview of the final analysis approach we used to assess each paper.

\myparagraph{Metrics:}
We assessed the use of PPP Theory with two metrics:

\begin{enumerate}
\item \emph{Construct Focus}: Which PPP Theory construct(s) did the paper focus on?
\item \emph{Relationship Prevalence:} Did the paper identify relationships between PPP Theory constructs?
\end{enumerate}
\noindent These metrics allow us to categorize what a study is \textit{about} and whether it \textit{considers} PPP Theory.

For the construct focus metric (item 1 above), we categorize each paper into one of the following three types:
\begin{enumerate}
    \item \textit{Process/Policy:} The paper observes or measures process/policy.
    For example, He \etal measures the library migration process in the Java ecosystem~\citep{HeLarge2021}.
    
    \item \textit{Product:} The paper observes or measures a product.
    For example, Shen \etal study root causes and symptoms of deep learning compiler bugs~\citep{shen_comprehensive_2021}.
    
    \item \textit{Both:} The paper considers both.
    For example, Di Grazia \etal measure the adoption and use of Python type annotations (process) \textit{and} the resulting statically-detectable type errors in Python projects (product)~\citep{di_grazia_evolution_2022}.

\end{enumerate}

For the relationship prevalence metric (item 2 above), we categorize each paper into one of the following three types:
\begin{enumerate}
    \item \textit{Silent:} The paper makes no mention of PPP Theory relationships.
    For example, Shen \etal~\citep{shen_comprehensive_2021} report the characteristics of deep learning compiler bugs but do not explicitly describe how software engineering processes or policies can cause these bugs or should be influenced by bugs.
    \item \textit{Descriptive:} The paper \emph{mentions} a relationship between PPP Theory constructs.
    For example, He \etal~\citep{HeLarge2021} measure the library migration process and describe the importance of this process with respect to Java software products, but do not directly measure this relationship.
    \item \textit{Experimental:} The paper \emph{measures} a relationship between PPP Theory constructs.
    For example, Di Grazia \etal~\citep{di_grazia_evolution_2022} measures the relationship between using type annotations and the resulting number of type errors.
\end{enumerate}

Also, we categorized each paper based on the ownership of the empirical data employed in the study: \textit{Public} (papers involving publicly accessible data), \textit{Private} (data not accessible/proprietary), and \textit{Both} (papers incorporating both private and public data).

\myparagraph{Analysis Process:}
Our raters followed a modified version of Keshav's ``First Pass'' to quickly assess the PPP-Theoretic content of selected papers~\citep{howtoreadpaper}.
Raters proceeded as follows:
\begin{itemize}
    \item Read title, abstract, introduction, and research questions.
    \item Read section headings.
    \item Read the findings --- this includes highlighted key results, the discussion, and the Conclusion section.
\end{itemize}
After performing this ``First Pass,'' raters categorized the construct focus and relationship prevalence metrics for the paper according to the descriptions mentioned above.

Each paper was evaluated by two raters.
Inter-rater agreement was measured using Cohen's Kappa score~\citep{cohen1960coefficient}, and all disagreements were settled through discussion.
Prior to settling disagreements, our process produced a Kappa score of \KappaScoreConstruct for the construct classifications, \KappaScoreRelationship for rating the PPP relationships in each work, and \KappaScoreDataSource for the Data accessibility. 

\section{Results}

\begin{figure}[ht]
  \centering
  \includegraphics[width=1.0 \linewidth]{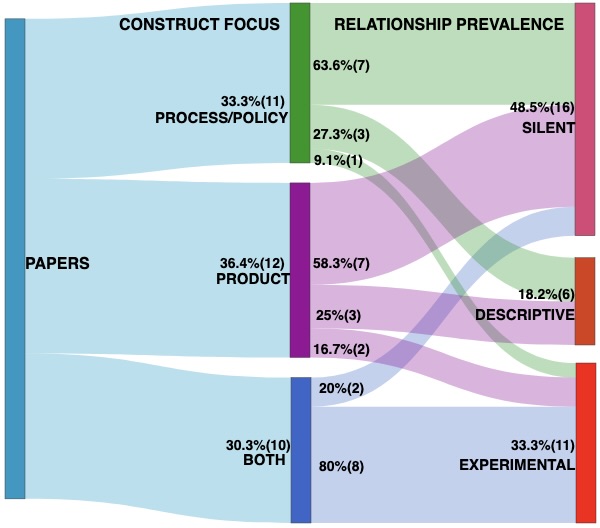}
  \caption{
  Distribution of analyzed papers based on Construct Focus (process/policy, products, both) and Relationship Prevalence (silent, descriptive, experimental).
  30\% of papers consider multiple constructs and 49\% of papers are silent about relationships between them.
  }
  \vspace{-0.15cm}
  \label{fig:sankey}
\end{figure}

In this section, we identify our results from assessing \RandomSamplePopulation empirical software engineering papers.
\cref{fig:sankey} summarizes our findings.

\subsection{Construct Focus}\label{result-construct}


The first stage of~\cref{fig:sankey} divides papers by their Construct focus.
Papers typically consider products (\ProductPercent) 
or policy/processes (\ProcessPolicyPercent).
The former observes the actions of software engineers and organizations, and the latter measures information about software artifacts.
\textbf{The smallest set considered both policy/process and products} (\BothPercent).

\subsection{Relationship Prevalence}

The second stage of~\cref{fig:sankey} divides papers by their relationship prevalence.
\textbf{Work experimenting with or measuring PPP Theory relationships was rare}: \ExperimentalPercent or \ExperimentalPapers of the papers.
About half of the papers (\SilentPapers papers or \SilentPercent) did not discuss relationships from the PPP Theory, and the remaining \DescriptivePapers papers (\DescriptivePercent) were descriptive.
As expected, considering the construct focus of a paper, the studies focused on a single construct tend to be silent or descriptive, and studies that consider both constructs tend to relate them.

\subsection{Other observations}
We considered the ownership of the data used in these works.
Most papers used public data (\PublicPercent), some private (\PrivatePercent), rarely both (\BothDataPercent).
We observed one trend:
of the four studies that used both data types, three experimentally showed a PPP relationship.
Perhaps studies with diverse data provide more insight into PPP relationships.

\section{Implications for Research}

We suggest four implications for the research community.

\vspace{0.08cm}
\textbf{(1) Incorporate PPP Theory into study designs:}
A surprising fraction of papers (\SilentPercent) did not consider these PPP relationships.
We do not wish to criticize these works; there is value in characterizing processes and in characterizing products, whether or not relationships are demonstrated between these constructs.
However, we wonder if the Empirical Software Engineering research community would benefit from a greater focus on the PPP-Theoretic basis for their measurements.
This was the original vision of empirical software engineering introduced in the 1980s and 1990s~\citep{brilliant1990analysis,basili_building_1999,perlis1981software}.
This could provide a meaningful way to address concerns about the interpretability and generalizability of our community's empirical research~\citep{kitchenham_preliminary_2002, siegmund_views_2015, gonzalez-barahona_reproducibility_2012, madeyski_would_2017}.
As a step towards this,
  the SIGSOFT Empirical Standards~\citep{sigsoft-empirical-standards} could be extended to provide guidance about
    epistemology (\emph{What is software engineering knowledge?}~\citep{SWEBOK-2023}),
    not just about
    methodology (\emph{How to obtain knowledge?}).
Some thoughtfulness about the PPP Theory could help authors analyze the Threats to the Validity of their work, without resorting to vague statements about generalizability in ``other contexts''.

\vspace{0.08cm}
\textbf{(2) Study the feedback relationship:}
Among the papers that did consider a relationship between policy/process and product, we note that there was a bias toward measuring the ``forward'' direction of~\cref{fig: PPP model}.
In our sample it is unusual for researchers to characterize and measure the role of feedback in the engineering process.
Although many works have observed the opportunity for failures to inform future engineering approaches~\citep{anandayuvaraj2022reflecting,amusuo_reflections_2022}, this appears to still be a gap in the literature.

\vspace{0.08cm}
\textbf{(3) Study more feed-forward relationships:}
Although the ``forward'' direction of relationships was more commonly examined, there are classes of constructs whose relationships were not examined in our sample.
Papers in this category considered topics like software organizational structure, software evolution, and software maintenance.
We did not see any papers on topics such as the effect of regulations (\eg GDPR), cybersecurity policies (\eg NIST 8397), or industry standards (\eg MISRA).
Greater industry collaboration might facilitate the study of such relationships.
Empirical software engineering research often examines open-source software --- those engineers lack the liability that motivates organizations to promote such policies and processes.

\vspace{0.08cm}
\textbf{(4) Disentangle Policy and Process:}
Lastly, the PPP Theory predicts separate roles of policy and process.
Policy interfaces with organizational goals, while process interfaces with the engineered product.
These constructs are generally entangled in the empirical software engineering literature, so in our model, we combined them in ~\cref{fig:method}.
Considering them as separate constructs may help the community develop a richer theory of software engineering.

\section{Threats to Validity}

\textbf{Construct:}
We rely on constructs and relationships defined by the PPP Theory.
Our specific operationalizations were derived from literature (\cref{sec: Background}), but distinguishing these can be difficult because they are often entangled.
We addressed this concern through interrater agreement,  achieving reasonable Kappa scores of \KappaScoreConstruct for PPP constructs and \KappaScoreRelationship for relationships between constructs.

\vspace{0.08cm}
\noindent
\textbf{Internal:}
We make no claims of cause and effect.

\vspace{0.08cm}
\noindent
\textbf{External:}
We sampled half of the empirical works from ICSE, ESEC/FSE, and ASE 2021-2022.
A longer time span or alternative venues might affect our results.
Based on our understanding of the recent research literature, we do not think time is a crucial variable.
These three venues are large general venues, so considering other venues seems unlikely to substantially shift our result.

\section{Conclusion}
In this paper, we have investigated the degree to which current empirical software engineering works consider the relationship between policies, processes, and software products (PPP relationship). We have reviewed {\RandomSamplePopulation} published works. Our results show that:
(1) Most empirical software engineering works are focussed on single constructs of the PPP theory model, and
(2) \DescriptivePercent of the reviewed works provided a description for the PPP relationships, while \SilentPercent of the works were silent on this PPP relationship.

Consequent to our results, we have made 4 suggestions to the software engineering research community on the significance of adopting the PPP Theory in future empirical studies. Our recommendations center on the need to further study the PPP theory constructs, and their forward and feedback relations. 

\section*{Data Availability}
Data is available on Zenodo: \url{https://doi.org/10.5281/zenodo.8277429}.

\ifANONYMOUS
\else
\section*{Acknowledgments}
We thank A. Kazerouni and the reviewers for their feedback.
We acknowledge financial support from
  NSF awards IIS-1813935, 
             SaTC-2135156, 
             and
             POSE-2229703, 
 as well as
  Cisco 
  and
  Rolls Royce. 
\fi

\balance
\clearpage

\bibliographystyle{IEEEtran}
\bibliography{references}

%
\raggedbottom
\pagebreak
\balance

\end{document}